\documentclass[showpacs,amssymb,preprintnumbers,nofootinbib,superscriptaddress]{revtex4}

\usepackage{amsmath}
\usepackage{graphicx}
\usepackage{epsfig}
\usepackage{latexsym}
\usepackage{amsfonts}
\usepackage{url,hyperref}
\usepackage{bm}

\begin{document}

\begin{figure}[t]
\vspace{-1.4cm}
\hspace{-16.25cm}
\scalebox{0.09}{\includegraphics{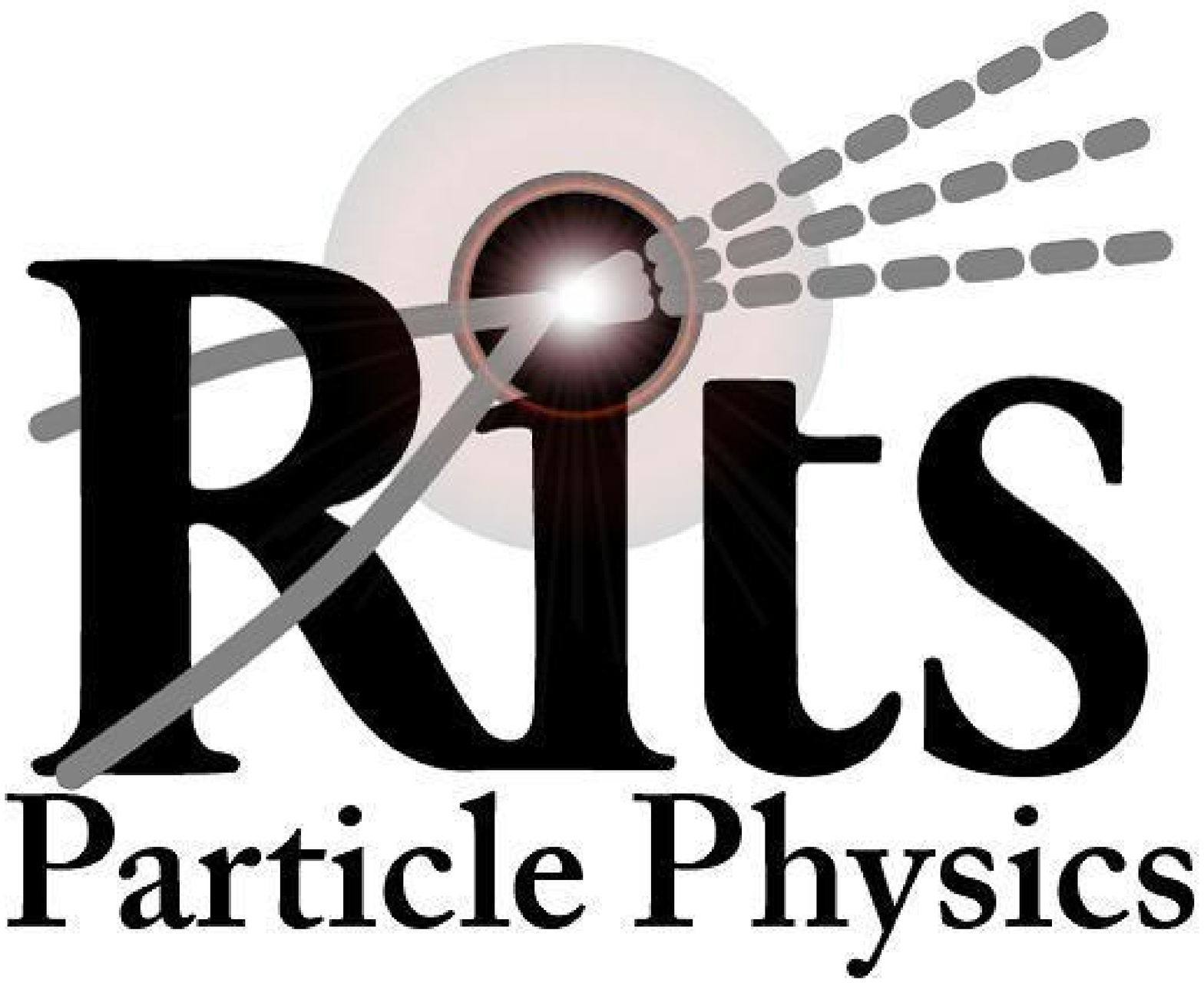}}
\end{figure}

\newcommand{\vp}{\varphi}
\newcommand{\nn}{\nonumber\\}
\newcommand{\beq}{\begin{equation}}
\newcommand{\eeq}{\end{equation}}
\newcommand{\bed}{\begin{displaymath}}
\newcommand{\eed}{\end{displaymath}}
\def\bea{\begin{eqnarray}}
\def\eea{\end{eqnarray}}
\newcommand{\veps}{\varepsilon}

\title{Bulk dominated fermion emission on a Schwarzschild background}

\author{H.~T.~Cho}
\email{htcho"at"mail.tku.edu.tw}
\affiliation{Department of Physics, Tamkang University, Tamsui, Taipei, Taiwan, Republic of China}
\author{A.~S.~Cornell}
\email[Email: ]{cornell@ipnl.in2p3.fr}
\affiliation{Universit\'e de Lyon, Villeurbanne, F-69622, France;
Universit\'e Lyon 1, Institut de Physique Nucl\'eaire de Lyon}
\author{Jason~Doukas}
\email[Email: ]{j.doukas"at"physics.unimelb.edu.au}
\affiliation{School of Physics, University of Melbourne, Parkville, Victoria 3010, Australia.}
\author{Wade~Naylor}
\email[Email: ]{naylor"at"se.ritsumei.ac.jp}
\affiliation{Department of Physics, Ritsumeikan University, Kusatsu, Shiga 525-8577, Japan}

\begin{abstract}
Using the WKBJ approximation, and the Unruh method, we obtain semi-analytic expressions for the absorption probability (in all energy regimes) for Dirac fermions on a higher dimensional Schwarzschild background. We present an analytic expression relating the absorption probability to the absorption cross-section, and then use these results to plot the emission rates to third order in the WKBJ approximation. The set-up we use is sufficiently general such that it could also easily be applied to any spherically symmetric background in $d$-dimensions. Our results lead to the interesting conclusion that for $d>5$ bulk fermion emission dominates brane localised emission. This is an example contrary to the conjecture that black holes radiate mainly on the brane.
\end{abstract}

\pacs{02.30Gp, 03.65Ge, 0470.Dy, 11.10.Kk}
\preprint{LYCEN 2007-13}
\preprint{RITS-PP-013}
\date{\today}
\maketitle

\section{Introduction}

\par In the last decade a great deal of attention has been focused on large extra-dimensional scenarios \cite{ADD}, where the hierarchy problem can be shifted into a problem of the scale of the extra-dimensions. These scenarios have also led to the somewhat striking prediction that black holes (BHs) may be observed at particle accelerators such as the LHC \cite{BHacc}. However, one poignant problem is that in order to suppress a rapid proton decay quarks and leptons need to be physically separated in the higher dimension(s). Such models are generically called split fermion models, see reference \cite{Split}, for example. Note that in supersymmetric versions of this idea the localizing scalars and bulk gauge fields will also have fermionic bulk superpartners. In this respect it is important to consider the properties of bulk fermions, something which until now has been largely overlooked. For a treatment of brane localized Hawking emission in the case of static BHs see reference \cite{HiHawk,Kanti}, for rotating BHs, see reference \cite{Rotate} as well as references \cite{BraneFerm}.

\par In a previous work we applied conformal methods, which allowed us to separate the Dirac equation on a higher dimensional spherically symmetric background, to discuss the quasinormal modes (QNMs) for Schwarzschild BHs\cite{CCDN}; where such a method was previously used in reference \cite{Das} to calculate low energy $s$-wave absorption cross-sections. In this work we shall use this same method to calculate the greybody factors \cite{Das} and emission rates for Dirac perturbations on a $d$-dimensional Schwartzschild background. Our results may also be of some interest to the {\it tense}-less limit of the tense branes in six dimensions discussed in reference \cite{Dai}.

\par Here we shall write the $d$-dimensional Schwarzschild background as:
\begin{equation}
ds^{2}=-f(r)dt^{2}+f^{-1}(r)dr^{2}+r^{2}d\Omega^{2}_{d-2} \,\,\, ,
\end{equation}
with
\begin{equation}
f(r)=1-\left(\frac{r_{H}}{r}\right)^{d-3} \,\,\, ,
\end{equation}
where the horizon is at $r=r_{H}$ and:
\begin{equation}
r_{H}^{d-3}=\frac{8\pi M\Gamma((d-1)/2)}{\pi^{(d-1)/2}(d-2)} \,\,\, .
\label{horiz}
\end{equation}
In reference \cite{CCDN} we used a conformal transformation of the metric to  separate the Dirac equation into a time-radial part and a $(d-2)$-sphere. Moreover, the radial part was reduced to a Schr\"odinger-like equation in the tortoise coordinate $r_{*}$:
\beq
\left(-\frac{d^2}{dr_{*}^{2}}+V_{1}\right)G=E^{2}G \,\,\, ,
\label{gequation}
\eeq
where $dr=f(r) dr_{*}$, and the potential is given by:
\begin{eqnarray}
V_1(r) &=&\kappa^2 {f\over r^2}+\kappa
f\frac{d}{dr}\left[\frac{\sqrt{f}}{r}\right] \,\,\, ,
\end{eqnarray}
where
\beq
\kappa = \ell + \frac{d-2}{2} = \frac{d}{2} - 1 \,\,\, , \,\,\, \frac{d}{2} \,\,\, , \,\,\, \frac{d}{2} + 1 \,\,\, , \dots
\eeq
It may be worth mentioning that the above potential reduces to the brane-localized results found in reference \cite{JP} when we set $\kappa=\ell+1$ and therefore provides an alternate derivation of the brane localized potential.

\par In the next section, we shall compare the absorption probability for this potential as calculated under various approximation techniques, namely, the low energy WKBJ, the first to third order intermediate WKBJ and the low energy Unruh method. Then we shall relate these quantities to the absorption cross-section and calculate the Hawking emission rate. After this we will compare bulk emission with brane emission (to third order in WKBJ) for various dimensions, before presenting our concluding remarks in section IV.

\section{Absorption probabilities via the WKBJ approximation}

\par In a recent work by two of the authors \cite{Cornell} we applied the intermediate WKBJ approximation (up to first order) to evaluate the absorption probability of a graviton to a static BH. The WKBJ approximation can, however, be applied at all energies (including low energy) as has been discussed in reference \cite{ChoWKB}. In Appendix A we calculate a different low energy approximation based on the method developed by Unruh \cite{Unruh}. This `{\it Unruh method}' leads to useful analytic results, but does not work well for high energy regimes.

\subsection*{Low Energy WKBJ}

\par In terms of the WKBJ approximation, in general, it will be convenient to make a change of variables to $x=Er$ \cite{Cornell}. This leads to the following form for the potential:
\beq Q(x_*)= 1 - \kappa^2 {f\over x^2} - \kappa f \frac{d}{dx}
\left[ \frac{\sqrt{f}}{x}\right] \,, \qquad \quad f(x) = 1 -
\left( \veps\over x\right)^{d-3},\label{definitionofq} \eeq where
\beq E^2 Q(x_*) = E^2 - V_1 \,\,\, , \eeq and $\veps = r_H
E$.\footnote{Note that in reference \cite{Cornell} we defined the
horizon, $r_H$, in terms of $\veps=r_H E^{d-3}$, however, by using
this different parameterisation here, we may compare this result
with the low energy Unruh approach.} As such, the Schr\"{o}dinger
equation, equation (\ref{gequation}), takes the form: \beq
\left(\frac{d^2}{dx_{*}^{2}}+Q\right)G=0 \,\,\, . \eeq

\par The low
energy absorption probability corresponds to the probability for a
particle to tunnel through the potential barrier (the barrier
penetration probability) and can be found in many standard quantum
mechanics text books (a derivation can also be found in reference
\cite{ChoWKB}). The result to first order in the low energy WKBJ
approximation is given by:
\begin{equation}
|{\cal A}_\kappa(E)|^2
=\exp\left[{-2\int^{x_2}_{x_1}\frac{dx'}{f(x')} \sqrt{-Q(x')}
}\right] \,\,\, ,
 \label{LEabprob}
\end{equation}
where $x_1$ and $x_2$ are the turning points, $Q(x_{1,2})=0$ or
$V_{1}(x_{1,2})=E^{2}$, for a given energy $E$ with potential
$V_1$. This approximation is valid for $V_1 \gtrsim E^2$ and as
long as we can solve for the turning points in $V_1(x)=E^{2}$.
Note that we can numerically integrate equation (\ref{LEabprob})
for each energy $E$ to obtain the absorption probability as a
function of $E$.

\subsection*{Intermediate  Energy: 3rd Order WKBJ}

\par As discussed in reference \cite{IW}, an adapted form of the WKBJ method can be employed to find the QNMs, or the absorption probability (which we are primarily interested in here), when the scattering takes place near the top of the potential barrier.
In the following we shall use the same notation as reference \cite{WG}, where we have confirmed their results to fourth order. However, for the purposes of this paper, we shall consider only up to and including third order, in which case we express the absorption probability as:
\beq
|{\cal A}_\kappa(E)|^2 = {1\over 1+e^{2S(E)}} \,\,\, ,
\label{IMabprob}
\eeq
where
\bea
S(E) &=& \pi k^{1/2} \left[ \frac{1}{2} z_0^2 + \left( \frac{15}{64} b_3^2 - \frac{3}{16} b_4 \right) z_0^4 +\left( \frac{1155 }{1024}b_3^4-\frac{459 }{128}b_4 b_3^2+\frac{95}{32}b_5 b_3+\frac{67}{64}b_4^2-\frac{25 }{16}{b_6}\right) z_0^6  \right] \nn
& & + \pi k^{-1/2} \left(\frac{3}{16} b_4 - \frac{7}{64} b_3^2\right) - \pi k^{-1/2}
\left( \frac{1365}{2048} b_3^4 -\frac{525}{256} b_3^2 b_4+\frac{95}{64}{b_5} {b_3} + \frac{85}{128}b_4^2 -\frac{25}{32}{b_6} \right)z_0^2 \nn
& & + {\cal O} (k^{-3/2} z_0^0,\, k^{-1/2}z_0^4, \,k^{1/2}  z_0^8) \,\,\, .
\eea
Note that the first order result comprises of just the first term in this expression, while the second order result consists of the second term on the first line and the first term on the second line.

\par We would like to draw the readers attention to the fact that as we go to higher order the approximation becomes valid for lower energies. However, as can be seen from figure \ref{abs}, even orders in the intermediate WKBJ method drop back down to zero for large energy. For this reason we shall work to third and not fourth order in our calculations, as odd orders have the nice property that $|A|^2 \to 1$ for large energy, making numerical work with these orders easier. Also, note that the first order WKBJ under-predicts the absorption probability for $\veps \sim {\cal O}(1)$ as compared to the second order WKBJ result. We have compared these results up to fourth order in the WKBJ approximation \cite{WG}, and the essential features of the energy absorption profile (such as dipping back down to zero for even orders) do not change when going from second to fourth order; nor do features of the first order change when going to third order, see figure \ref{abs}.

In this method $Q$ in equation (\ref{definitionofq}) is expanded
near the peak ($x=x_0$), in terms of a new variable $z=x-x_0$:
\beq Q(z)=k \left(z^2-z^2_0+\sum_{n=3}b_n z^n\right) \,\,\, , \eeq
where \beq z_0^2 \equiv - 2 \frac{Q_{0}}{Q_0^{\prime\prime}}
\,\,\, ,\qquad \qquad k \equiv \frac{1}{2} Q_0^{\prime\prime}
\,\,\, , \qquad \qquad b_n \equiv \frac{2}{n! Q_0^{\prime\prime}}
\left. \frac{d^n Q}{dr^n} \right|_0 \,\,\, . \eeq Note that the
subscript $0$ represents the maximum of $Q$, and that for our
purposes it is more convenient to work in terms of $z$, rather
than $z_*$. However, given that the WKBJ form of the potential is
in terms of the parameter $z_*$, including its derivatives, we can
convert derivatives in terms of $z$ into $z_*$ via the equation
$dz_*/dz = 1/f(z)$. See reference \cite{CCDN} for a discussion of
how to locate the maximum.

\subsection*{High Energy}

\par For high energies the absorption probability tends to unity, and the cross-section reduces to that of the classical cross-section, see reference \cite{Cornell}. However, as discussed in reference \cite{ChoWKB}, there will always be small corrections to the large energy limit. A high energy WKBJ approach can be applied in this limit, but for the purposes of this current study it will be sufficient to use $|{\cal A}_\kappa(E)|^2=1$ (given that it reproduces the high energy geometric optics limit, see references \cite{Kanti,Cornell}).

\subsection*{Absorption Probability Results}

\begin{figure}[t]
\begin{center}
\vskip -0.1cm
\epsfig{file=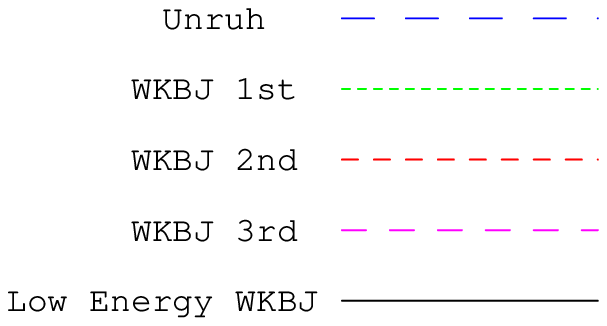,width=.25\textwidth}\hskip -0.5cm
\epsfig{file=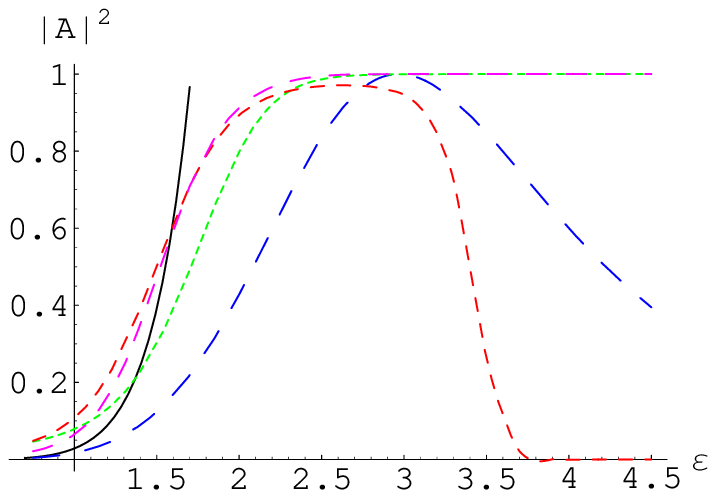,width=.39\textwidth}  \hskip -0.5cm
\epsfig{file=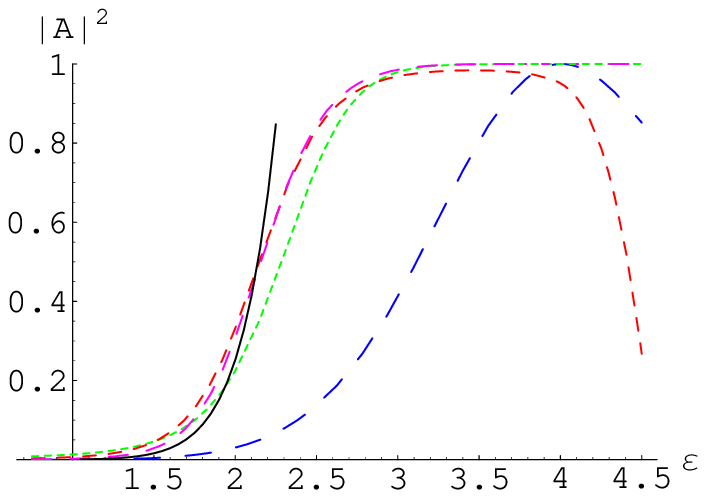,width=.39\textwidth}
\end{center}
\vspace{-.5cm}
\caption{{\it Plots of the absorption probability, via various approximation schemes, for $d = 7$ and the first two angular momentum channels: $\ell=0$ (left) and $\ell=1$ (right). Note the {\it Unruh} result, equation (\ref{Unruh}), is strictly speaking only valid for $\veps \ll 1$}.}
\label{abs}
\end{figure}
\par The results of these analyses have been shown in figure \ref{abs}. In this figure we presented plots for the first, second and third order WKBJ Iyer and Will method \cite{IW}, which is an intermediate WKBJ approximation, as well as the low energy WKBJ result (to first order) and the analytical method of Unruh \cite{Unruh}. The WKBJ approximation, in general, is accurate for larger angular momentum channels, whereas the Unruh approach is valid for only the lowest angular momentum channels and $\veps \ll 1$ (namely small BHs). However, it is interesting to note that although the low energy WKBJ result does not agree exactly with the Unruh result, they both tend to zero for $\veps \to 0$. On the other hand, unlike the Unruh result, the low energy WKBJ is valid for energies up to $\veps \sim {\cal O}(1)$, where it matches onto the intermediate WKBJ.\footnote{In order for the first order low energy WBKJ to match onto the Iyer and Will second/third order result we are really required to evaluate it to second/third order as well.}

\section{Emission Rates}
\par The emission rate for a massless fermion from a BH is related to the cross-section by a $d^{d-1}k$ dimensional momentum integral times a fermionic thermal temperature distribution:
\beq
{d {\cal E}\over dt} = \sum_{\lambda,E} \sigma_{\lambda,E} {E \over e^{E \over T_H}+1} {d^{d-1}k \over (2\pi)^{d-1}} \,\,\, , \label{EmX}
\eeq
where $T_H$ is the Hawking temperature, $\sigma_{\lambda,E}$ are the greybody factors and the sum is a generic sum over all angular momentum and momentum variables. In Appendix B using the method of Cardoso {\it et al.} \cite{Cardoso} we relate the greybody factor to the absorbtion probability:
\beq
\sigma_{\lambda,E} = {1\over 2 \Omega_{d-2}} \left( 2\pi\over E\right)^{d-2} \sum_\kappa D_\kappa
|{\cal A}_\kappa(E)|^2 \,\,\, . \label{Xsec}
\eeq
Given that angular integration over the momentum for a massless field ($|k|=E$) leads to the Jacobian $\int d^{d-1}k=\int \Omega_{d-2} E^{d-2} dE$, the fermion emission rate can be expressed solely in terms of the absorption probability:
\beq
{d {\cal E}\over dt} = \sum_{\kappa} \int {dE \over 2\pi} {E \over e^{E \over T_H}+1}
D_\kappa |{\cal A}_{\kappa}(E)|^2 \,\,\, . \label{Emission}
\eeq
In the above we have used $D_\kappa$ as the degeneracy (as defined in equation (\ref{degeneracy})) and the sum over $\kappa$ for $\kappa = \pm \left( \frac{d}{2}-1 \right)$, $\pm \frac{d}{2}$, $\pm \left(\frac{d}{2} + 1 \right)$, $\dots$. However, since the integrand depends only on the absolute value of $\kappa$ we only sum for $\kappa \geq 0$ and multiply by a factor of two. We then recover a result identical to that for a scalar field \cite{Kanti}, except for a difference in sign due to fermion statistics.

\par After changing variables to $\varepsilon = Er_H$, and using the fact that the Hawking temperature is $T_H=(d-1)/(4\pi r_H)$, we obtain:
\beq {d^2{\cal E}\over dE dt} = {1 \over \pi r_H}
\sum_{\kappa>0}{\veps \over e^{4\pi \veps \over d-1} + 1}
D_{\kappa} |{\cal A}_{\kappa}(\veps)|^2 \,\,\, . \label{Miss} \eeq
As such, the evaluation of the emission rate is now a simple task,
as $|{\cal A}_{\kappa} (\veps)|^2$ can be obtained either
numerically or via the WKBJ method in each appropriate energy
regime. Note that we have presented an example of these emission
rates for various values of $d$ in figure \ref{emiss} (up to third
order WKBJ), these results shall be discussed in the conclusion.
Importantly, when we come to consider brane-localized emissions we
simply set $\kappa=\ell+1$ and $D_\kappa=2(\ell+1)$, on assuming
$\kappa\geq 0$. These are also plotted in figure \ref{emiss} for
various dimensions $d$.

\subsection*{Third Order WKBJ Emission Spectrum}

\par
\begin{figure}[h]
\begin{center}
\vskip -0.1cm \hskip -0.25cm
\epsfig{file=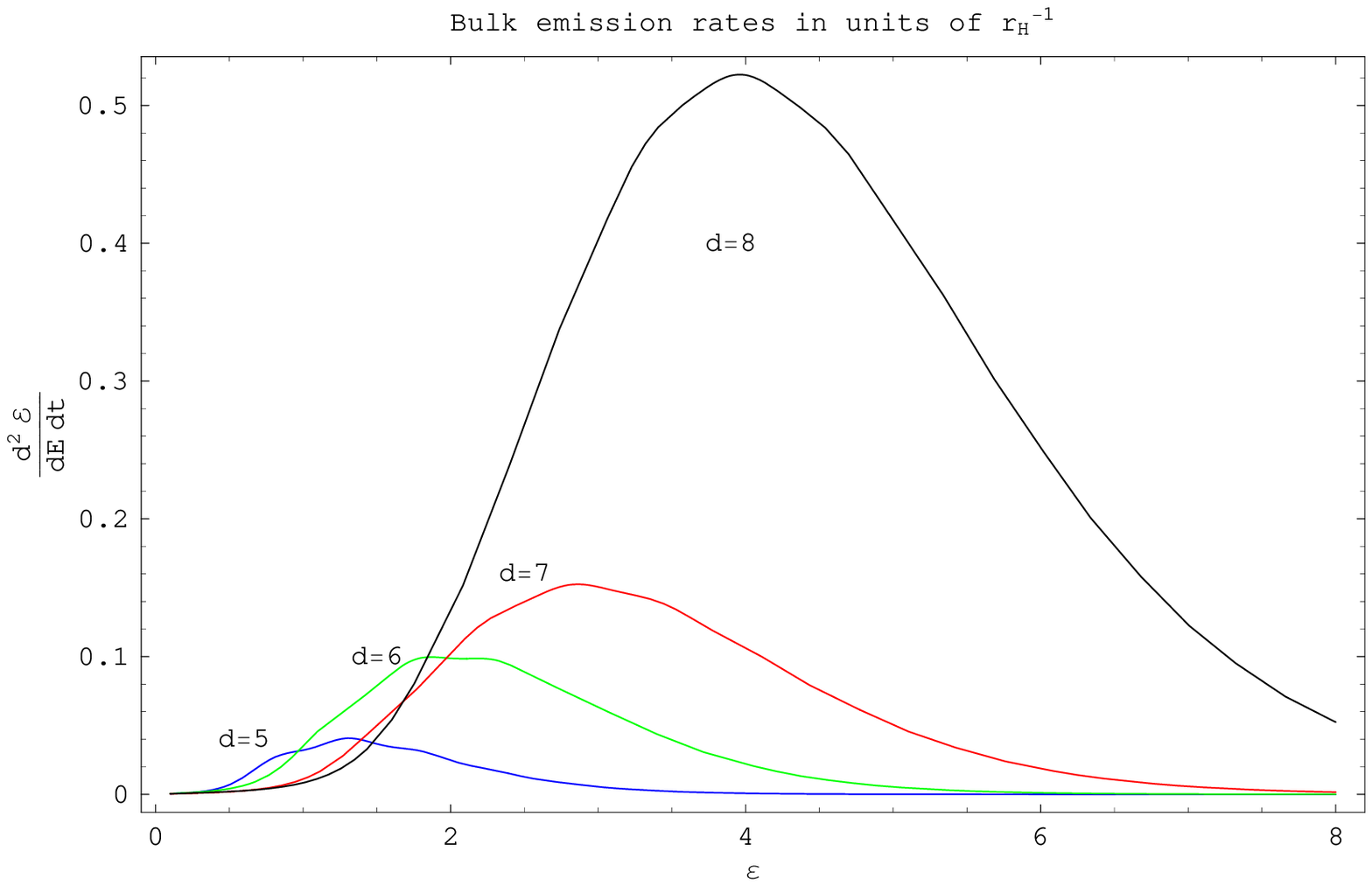,width=.5\textwidth}
\hskip -0.cm
\epsfig{file=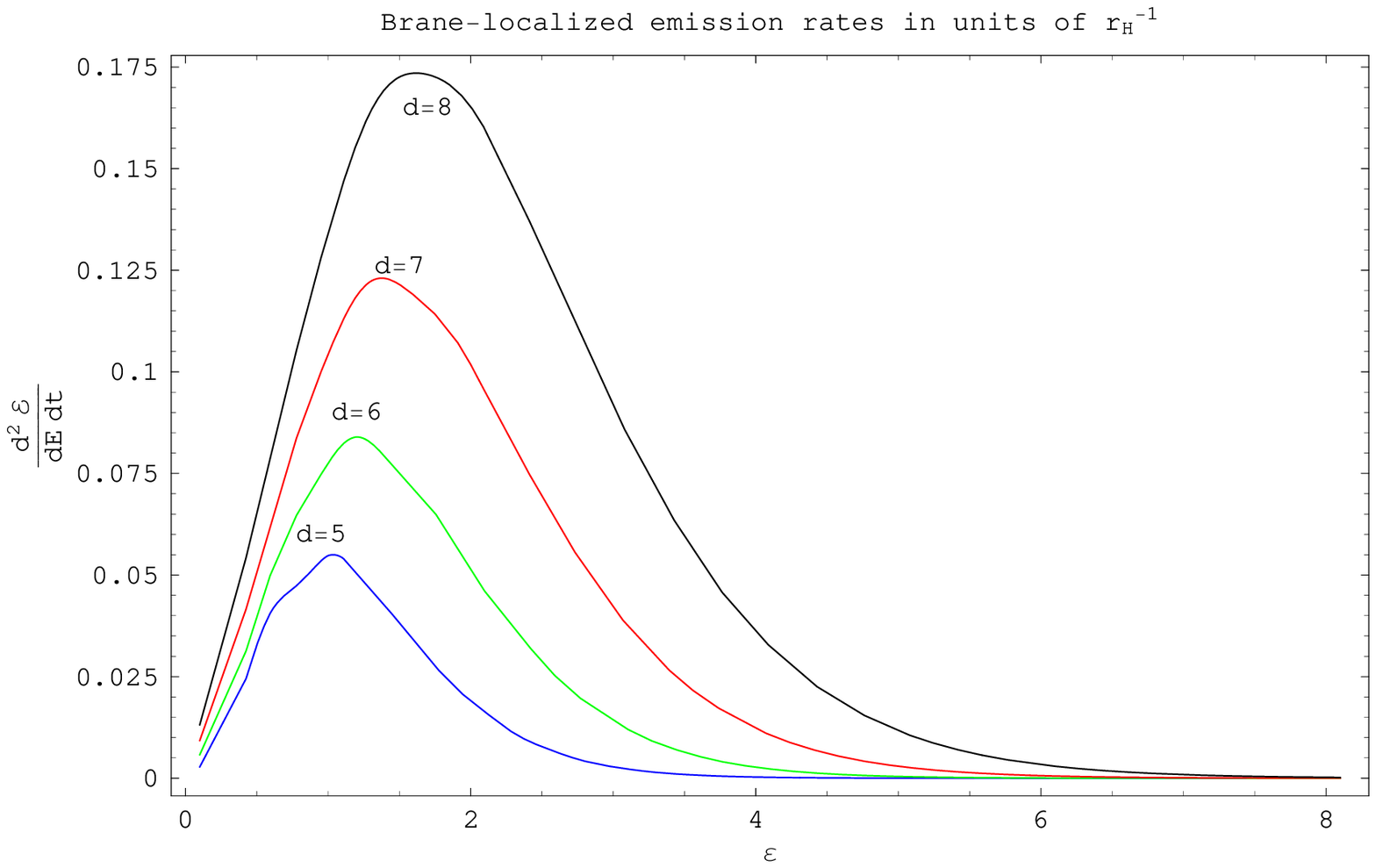,width=.5\textwidth}
\vspace{-.25cm}
\caption{{\it Plots of the fermion emission rates for $d = 5$, $6$, $7$ and $8$ using the third order intermediate energy WKBJ method, where we compare bulk emissions (left panel) with brane-localized emissions (right panel)}.}
\label{emiss}
\end{center}
\end{figure}

 Since the intermediate WKBJ is accurate for $\veps>1$ we plot the emission rates in this approximation, as shown in figure {\ref{emiss}}.

\par We have also calculated the total power by integrating over $\veps$, see equation (\ref{Miss}). The results are shown in Table \ref{tab:BulkBranePower}. From these results we find that for $d>5$ the emission is predominantly into the bulk.
\begin{table}[h]
\centering
\begin{tabular}{|c|c|c|c|c|c|c|}
\hline
Dimension $d$~~~~&5&6&7&8 & 9 & 10\\
\hline
$d{\cal E}_{\rm Bulk}/dt$ & 0.0579 &0.1771 &0.3380 &1.4731 &3.56403 &18.2606 \\
&&&&&&\\
\hline
$d{\cal E}_{\rm brane}/dt$ &0.0708 &0.1172 &0.204 &0.3435 &0.554892 &0.860165\\
&&&&&&\\
\hline
${d{\cal E}_{\rm Bulk}/ dt \over d{\cal E}_{\rm brane} / dt}$&0.8181&1.5109&1.6587&4.2880 &6.42292 &21.9019\\
&&&&&&\\
\hline
\end{tabular}
\caption{{\it A comparison of the bulk and brane-localised power spectrum up to $d=10$, where in equation (\ref{horiz}) we have changed units from $r_H$ to $M$ and set $M=1$}.}
\label{tab:BulkBranePower}
\end{table}
\par Note that in order to obtain convergence in equation (\ref{Miss}) we must choose some value of $\kappa_{max} > \veps$ and to ensure this we have taken $\kappa_{max} = 34+\tfrac{d}{2}$.

\par
\begin{figure}[h]
\hfill
\begin{minipage}[t]{.45\textwidth}
\begin{center}
\epsfig{file=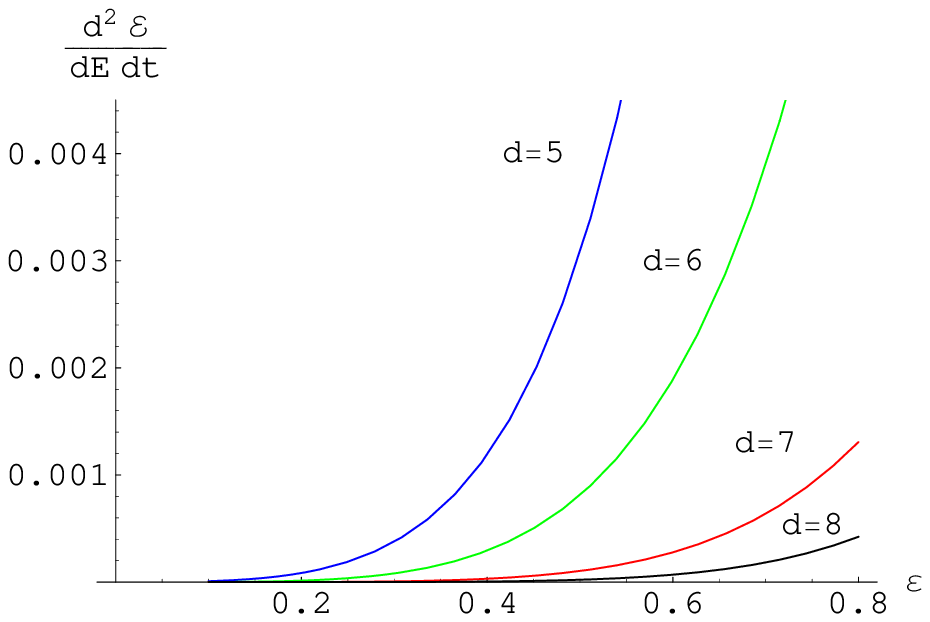, width=0.8\textwidth}
\end{center}
\end{minipage}
\hfill\begin{minipage}[t]{.45\textwidth}
\begin{center}
\epsfig{file=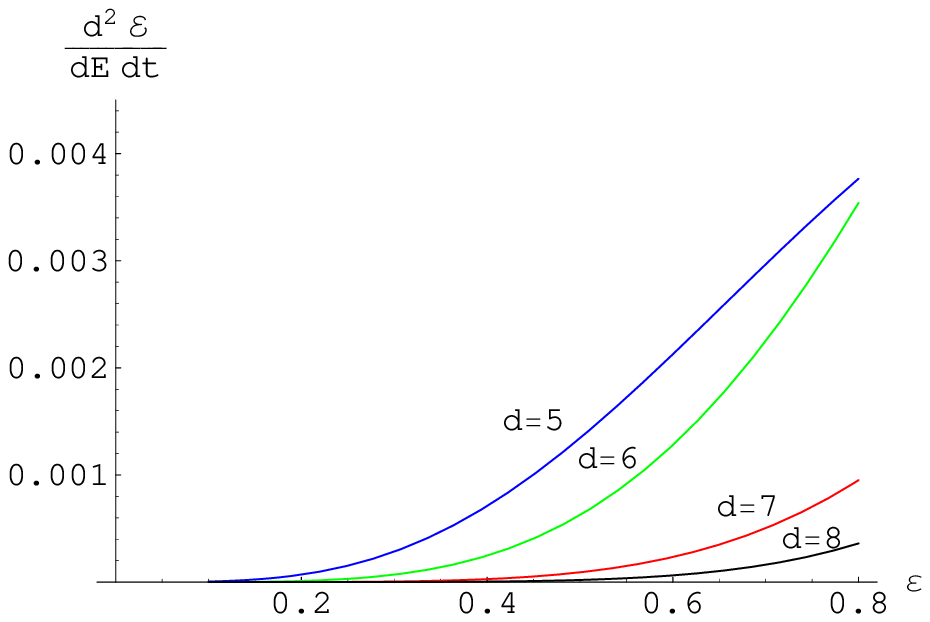, width=0.8\textwidth}
\end{center}
\end{minipage}
\caption{{\it Plots of the bulk fermion emission rates for $d = 5$, $6$, $7$ and $8$ using the low energy WKBJ method (left) and the Unruh method (right). The emission sums were found to converge for $\kappa_{\rm max} =\tfrac{d}{2}$. Note that the ordering of the lines is opposite to that in the high energy limit, verifying that the crossing of lines near $\veps \sim 1$ in figure \ref{emiss} is not a break down of the intermediate WKBJ approximation but a real effect for bulk fermions}.}
\label{LowUnruhEmission}
\end{figure}

 From our results we should note that of interest is the region where $\veps \sim 1$ in the bulk emission plots, where we see that the lines are crossing over. To be sure that this was not due to the semi-analytic approximation breaking down we computed the emission rates in the low energy regime using the low energy WKBJ and Unruh methods as shown in figure \ref{LowUnruhEmission}. From these plots we clearly see an opposite ordering of the lines to those from the high energy region in figure \ref{emiss}, verifying that a crossing does indeed take place.

\section{Concluding remarks}

\par In this paper we have presented new results for the emission rate of a massless Dirac field on a bulk $d$-dimensional Schwarzschild background, where the method we used in reference \cite{CCDN} is sufficiently general and could be applied to other spherically symmetric backgrounds (and also for massive bulk Dirac fields). The main result is that fermions are mainly emitted into the bulk for $d>5$, as we have shown, see Table \ref{tab:BulkBranePower}, which is in contrast to the scalar field case \cite{Kanti} and for bulk to brane photons \cite{Jung}. This is an example contrary to the conjecture that BHs radiate mainly on the brane \cite{EHM}. Furthermore, bulk dominated fermion emission is also consistent with the original motivation for split-fermions, namely that of a suppression of a rapid proton decay \cite{Split}.

\par We also highlighted how semi-analytic results can be obtained by considering different versions of the WKBJ approximation, where we also compared this to the low energy analytic results derived from the method first developed in reference \cite{Unruh}, see figure \ref{abs}. We also used the low energy WKBJ approximation, where it should be stressed that this approximation extends further than the range of the Unruh result (right up to intermediate energies) in the plots of $|A|^2$, again see figure \ref{abs}. However, in terms of the emission rates for low energy, see figure \ref{LowUnruhEmission}, the range of validity of $\veps$ extends up to $\veps \approx 0.5$, due to this energy range, for the emissions, being dominated by the two lowest angular momentum channels $\ell=d/2-1$ and $d/2$, essentially an $s$-wave scattering.

\par In figure \ref{emiss} we have plotted the third order WKBJ approximation of the emission rates for various $d$. As can be seen from this figure, in the bulk emission case, there is an interesting behaviour around $\veps \sim {\cal O}(1)$ where the emission rates in different dimensions cross. That is, at a certain intermediate energy there is a region where the emission rates become approximately independent of the dimension, $d$. We initially discovered this feature when plotting the first order WKBJ emission rates and found that this effect persisted to third order in the WKBJ approximation. The cross-over effect is confirmed by plotting the emission in the low energy region using the Unruh method, see figure \ref{LowUnruhEmission}, and plotting the geometric optics limit $|{\cal A}_\kappa(E)|^2=1$ at high energy and observing an opposite ordering of lines. We are unaware of this curious feature being reported anywhere else in the literature.

\par The results, when compared with the brane-localized fermion results of references \cite{BraneFerm,Kanti} reveal that bulk fermion modes result in much larger emission rates (though in both cases larger $d$ results in greater emission rates).  These results also agree qualitatively with our work for the QNMs on such a background \cite{CCDN}, where the BH damping rate was found to increase with dimension. Interestingly, the cross-over feature is absent from the emission rates of the brane-localized fermions, as can be seen in the right panel of figure \ref{emiss}, also see figure 4 of reference \cite{Kanti}.\footnote{Figure 4 of reference \cite{Kanti} is a Log plot not a Linear plot, however, there is no observable cross-over behaviour at intermediate energies.}  Thus, this feature appears to be specific to bulk species as we have also verified the same behaviour for bulk, but not brane-localized, scalar fields (this can also be observed in plots made in reference \cite{Kanti}).

\par The next step would be to investigate rotating solutions for bulk/split fermions (which have yet to be found for even a single rotation parameter)\footnote{See reference \cite{Kanti2} for scalar bulk and brane results for low energy and rotation  
parameter.}. However, in order to obtain these emissitivities the absorption cross-section would also need to be related to the absorption probability, but the method of Cardoso {\it et al.} \cite{Cardoso}, see Appendix B, does not work. That said, the energy flux could be derived by using the energy-momentum tensor for a spin-half field and comparing the flux at spatial infinity with that of the horizon (for a nice discussion of massless scalar fields see reference \cite{LWY} and the references therein). We are also currently applying our methods to investigate the effect of brane tension \cite{Dai} on our results. Finally, in a forthcoming work we intend to present results for $|A|^2$ up to 6th order in the intermediate WKBJ approximation, which follows along the lines of the work by Konoplya for QNMs \cite{Konoplya6th}, where we will compare it to exact numeric 
results.

\section*{Acknowledgments}

\par  HTC was supported in part by the National Science Council of the Republic of China under the Grant NSC 96-2112-M-032-006-MY3. JD wishes to thank Dr. G.~C.~Joshi for his advice and supervision during the production of this work. The authors would also like to thank Prof. M.~Sasaki for his interesting and useful discussions on this topic.

\appendix

\section{Low Energy and Momentum}

\par The case for the lowest angular momentum channel, $\kappa=d/2-1$, has been considered for bulk fermions in reference \cite{Das}. However, for completeness, we shall briefly discuss the low energy low angular momentum limit for general $\kappa$, which is valid for small BHs. Here we shall follow the method of Unruh \cite{Unruh}, but as the generalization to bulk fermions follows almost identically to the brane-localized case discussed in reference \cite{JP} we shall only briefly highlight the steps (referring the reader to reference \cite{JP} for a fuller discussion). Finally, note that in the following we shall work in terms of the potential $V_2$, not $V_1$, see reference \cite{CCDN}.

\par In terms of the $G$ component in equation (21) of reference \cite{CCDN}, the near horizon limit (which corresponds to $r\to r_H$ and $f(r)\to 0$) becomes:
\beq
\frac{d^2 G_{NH}}{d r_*^2} + E^2 G_{NH} = 0 \,\,\, ,
\eeq
where some simple deliberation leads to the following outgoing wave solution:
\beq
G_{NH} = A_{I} e^{-i E r_*} \approx A_{I} e^{-i \frac{\veps r_H}{d-3} \ln f} \,\,\, .
\eeq
Next, in the intermediate region, we assume that $E^2$ is much smaller than the other terms and equation (21) of reference \cite{CCDN} becomes (using the potential $V_2$):
\beq
\frac{d^2 G_{IM}}{d r_*^2} - \left( \kappa^2 {f\over r^2} - \kappa f \frac{d}{dr} \left[ \frac{\sqrt{f}}{r} \right] \right) G_{IM} = 0 \,\,\, . \label{2ndIM}
\eeq
Following the approach of Unruh \cite{Unruh} and defining:
\beq
H_{IM} \equiv \frac{d G_{IM}}{d r_*} + \frac{\kappa \sqrt{f}}{r} G_{IM} \,\,\, , \label{1storder}
\eeq
we can transform equation (\ref{2ndIM}) into a first-order differential equation of the form:
\beq
\frac{d H_{IM}}{d r} - \frac{\kappa}{\sqrt{f} r} H_{IM} = 0 \,\,\, . \label{HIM}
\eeq
The solution of equation (\ref{HIM}) is then:
\beq
H_{IM} = B_{II} \left( \frac{1 - \sqrt{f}}{1 + \sqrt{f}} \right)^{-\frac{\kappa}{d-3}} \,\,\, .
\eeq
Inserting the above result back into equation (\ref{HIM}) leads to \cite{Unruh,JP}:
\beq
G_{IM} = A_{II} \left( \frac{1 - \sqrt{f}}{1 + \sqrt{f}} \right)^{-\frac{\kappa}{d-3}} + B_{II} \cal{G}_{IM} \,\,\, ,
\eeq
where $\cal{G}_{IM}$ is a particular solution of:
\beq
\frac{d \cal{G}_{IM}}{d r} + \frac{\kappa}{\sqrt{f} r} {\cal{G}_{IM}} = \frac{1}{f} \left( \frac{1 - \sqrt{f}}{1 + \sqrt{f}} \right)^{\kappa\over{d-3}} \,\,\, .
\eeq
In general we have different solutions for positive or negative $\kappa$, and in what follows we shall just discuss the $\kappa < 0$ solutions. As such, for $\kappa < 0$ the solution to $\cal{G}_{IM}$ is:
\beq
{\cal{G}_{IM}} = r_{H} \left( \frac{1 - \sqrt{f}}{1 + \sqrt{f}} \right)^{-\frac{\kappa}{d-3}} \int_1^{\sqrt{f}} \frac{2}{(d-3) \rho} \frac{(1 - \rho)^{\frac{2 |\kappa| - (d-2)}{d-3}}} {(1 + \rho)^{\frac{2 |\kappa| + (d-2)}{d-3}}} d \rho \,\,\, ,
\eeq
with a slightly different solution for $\kappa>0$, see reference \cite{JP}.

\par To determine the flux at infinity we also require the solution in the far field region, and given that in this region $r_* \approx r$, or $r_* \to r$ (see equation (21) of reference \cite{CCDN}), leads to the simplification:
\beq
\frac{d^2 G_{FF}}{d r^2} + \left[E^2 - \frac{\kappa(\kappa+1)}{r^2} \right] G_{FF} = 0 \,\,\, . \label{farf}
\eeq
This solution can be expressed in terms of Bessel functions:
\beq
G_{FF}(r) = A_{III} \sqrt{\frac{r}{r_H}} J_{|\kappa + \frac{1}{2}|} (E  r) +  B_{III} \sqrt{\frac{r}{r_H}} Y_{|\kappa + \frac{1}{2}|} (E r) \,\,\, . \label{FF}
\eeq
By the large argument expansion of Bessel functions we can formally express the absorption probability as:
\beq
|{\cal A}_\kappa(E)|^2 = \frac{2 i \left(\frac{B_{III}}{A_{III}} - \frac{B_{III}^{\ast}}{A_{III}^{\ast}} \right)} {\left|1 + i \frac{B_{III}}{A_{III}} \right|^2} \,\,\, .
\eeq
Finally, all that is needed is to find the ratio of $ B_{III}/A_{III}$, which is done by matching the far-field, intermediate and near horizon regions. This leads to (for details see references \cite{Unruh,JP}):
\beq
\frac{B_{III}}{A_{III}} = - \frac{i \pi}{2^{-\frac{4 \kappa}{d-3}} \Gamma^2 \left( \frac{1}{2} - \kappa \right)} \left( \frac{E  r_H}{2} \right)^{-2 \kappa}~, \qquad \quad \qquad \frac{B_{III}}{A_{III}} = \frac{\pi 2^{-\frac{4 \kappa}{d-3}}}{i \Gamma^2 \left(\kappa + \frac{1}{2} \right)} \left( \frac{E r_H}{2} \right)^{2 \kappa} \,\,\, ,
\eeq
for $\kappa<0$ and $\kappa > 0$ respectively \cite{Unruh,JP}. It turns out, therefore, that the absorption probability is independent of the sign of $\kappa$:
\beq
|{\cal A}_{\kappa}(E)|^2 = \frac{\left( \frac{4 \pi 2^{-\frac{4 |\kappa|}{d-3}} \left( \frac{E r_H}{2} \right)^{2 |\kappa|}}{ \Gamma^2 \left(|\kappa| + \frac{1}{2} \right)} \right)}{ \left[ 1 + \frac{\pi  2^{-\frac{4 |\kappa|}{d-3}} \left( \frac{E  r_H}{2} \right)^{2 |\kappa|}}{ \Gamma^2 \left(|\kappa| + \frac{1}{2} \right)} \right]^2} \approx \frac{4 \pi 2^{-\frac{4 |\kappa|}{d-3}} \left( \frac{E r_H}{2} \right)^{2 |\kappa|}}{ \Gamma^2 \left(|\kappa| + \frac{1}{2} \right)} \,\,\, .
\label{Unruh}
\eeq
When these results are substituted into the relation for the cross-section this leads to the bulk
generalization for the ratio between the bulk fermion and scalar cross-sections for general $\kappa$ \cite{JP}.\footnote{Given that this approximation is only valid for $\veps\ll 1$ it can be expanded in a power series to give the approximate result on the right.}

\section{Absorption Cross-section}

\par In this appendix we shall discuss the relationship between $|{\cal A}_\kappa(E)|^2$ and the total absorption cross-section $\sigma$. In what follows we shall follow the approach of Cardoso {\it et al.} \cite{Cardoso} by integrating the total absorption cross-section over the direction of the incident plane wave and then dividing the result by the total solid angle (recall that the spacetime is spherically symmetric). Note that with this method one need not know the exact expansion of the incident plane wave in terms of the spherical ones.

\par As is done in reference \cite{Cardoso}, we concentrate on the case at spatial infinity, that is, in a $(d-1)$-dimensional Euclidean space, using a plane spinor wave:
\begin{equation}
\psi_{\rm plane}^{(E,\hat{k},s)}(\vec{x}) = e^{-iEt}e^{iE\hat{k}\cdot\hat{x}} \psi_{0}^{(\hat{k},s)} \,\,\, .
\end{equation}
Here $E$ is the energy, $\hat{k}$ the direction, and $s=\pm$ is the helicity of the wave. Note that as $\psi_{0}^{(\hat{k},s)}$ is a constant spinor it satisfies the equation:
\begin{eqnarray}
\gamma^{\mu}\partial_{\mu}\psi_{\rm plane}^{(E,\hat{k},s)}=0\Rightarrow\gamma^{0}\psi_{0}^{(\hat{k},s)} - \hat{k}_{i}\gamma^{i}\psi_{0}^{(\hat{k},s)} = 0 \,\,\, . \label{constspinor}
\end{eqnarray}
Note that the flux of the spinor wave is given by $J^{\mu}=-\bar{\psi}\gamma^{\mu}\psi$, where the minus sign is due to $(\gamma^{0})^{2}=-1$. For the plane spinor wave:
\begin{eqnarray}
\hat{k}_{i}J^{i} =
-\hat{k}_{i}(\psi_{0}^{(\hat{k},s)})^{\dagger}e^{iEt}
e^{-iE\hat{k}\cdot\vec{x}}\gamma^{0}\gamma^{i}e^{-iEt}
e^{iE\hat{k}\cdot\vec{x}}\psi_{0}^{(\hat{k},s)} & = & 1 \,\,\, ,
\end{eqnarray}
where we have used equation (\ref{constspinor}) and have assumed that $\psi_{0}^{(\hat{k},s)}$ is normalized. This shows that the plane wave has unit flux along the $\hat{k}_{i}$ direction.

\par Next, consider a spherical ingoing spinor wave:
\begin{eqnarray}
\psi_{\rm spherical}^{(E,\lambda)}=\frac{i}{r^{(d-2)/2}}e^{-iEt}e^{-iEr} \left(
\begin{array}{c}
1/\sqrt{2} \\ -1/\sqrt{2}
\end{array}
\right) \otimes\chi_{\lambda} \,\,\, ,
\end{eqnarray}
where $\chi_{\lambda}$ is an orthonormal spinor on the unit $(d-2)$-sphere, that is:
\begin{equation}
\int d\Omega_{d-2}(\chi_{\lambda})^{\dagger}\chi_{\lambda'}=\delta_{\lambda\lambda'} \,\,\, .
\end{equation}
The radial flux of this ingoing spherical wave is given by:
\begin{eqnarray}
J^{r} & = & -(\psi_{\rm spherical}^{(E,\lambda)})^{\dagger}\gamma^{0} \gamma^{r}\psi_{\rm spherical}^{(E,\lambda)} \nonumber\\
& = & -\frac{1}{r^{d-2}}\left( \left(
\begin{array}{cc}
1/\sqrt{2} & -1/\sqrt{2}
\end{array}
\right)(-i\sigma^{3})(\sigma^{2}) \left(
\begin{array}{c}
1/\sqrt{2} \\ -1/\sqrt{2}
\end{array}
\right) \right)(\chi_{\lambda})^{\dagger}\chi_{\lambda} \nonumber\\
& = & -\frac{1}{r^{d-2}}(\chi_{\lambda})^{\dagger}\chi_{\lambda} \,\,\, .
\end{eqnarray}
In which case:
\begin{equation}
N = -r^{d-2}\int d\Omega_{d-2}J^{r}=1 \,\,\, .
\end{equation}
Note that this spherical wave represents one ingoing particle per unit time. Following Cardoso {\it et al.}, we define $\alpha(E,\hat{k},s,\lambda)$ such that:
\begin{equation}
\psi_{\rm plane}^{(E,\hat{k},s)}=\sum_{\lambda}\alpha(E,\hat{k},s,\lambda)
\psi_{\rm spherical}^{(E,\lambda)}+{\rm outgoing\ part} \,\,\, .
\end{equation}
The total absorption cross-section for this particular incident plane wave is given by:
\begin{equation}
\sigma^{(E,\hat{k},s)} = \sum_{\lambda}|\alpha(E,\hat{k},s,\lambda)|^{2}
|{\cal A}^{s=1/2}(E,\lambda)|^{2} \,\,\, ,
\end{equation}
where $|{\cal A}^{s=1/2}(E,\lambda)|^{2}$ is the transmission probability.

\par Since the BH is spherically symmetric ${\cal A}^{s=1/2}(E,\lambda)$ is independent of $\hat{k}$ and $s$. Therefore, one can sum over $s$ and integrate over $\hat{k}$, and then divide the result by 2 times the total solid angle, to obtain the same absorption cross-section. That is:
\begin{eqnarray}
\sigma & = & \frac{1}{2\Omega_{d-2}}\sum_{s=\pm}\int d\Omega_{d-2} \sigma^{(E,\hat{k},s)}\nonumber\\
& = & \sum_{\lambda}\left(\frac{1}{2\Omega_{d-2}}\sum_{s=\pm} \int d\Omega_{d-2} | \alpha(E,\hat{k},s,\lambda) |^{2} \right)|{\cal A}^{s=1/2}(E,\lambda)|^{2}\label{sigma} \,\,\, .
\end{eqnarray}

\par Next, we evaluate the quantity in the bracket. To do so we first consider the projection of the plane wave onto the spherical ones with different energies. That is, we integrate over the Euclidean $(d-1)$-dimensional space:
\begin{eqnarray}
\int d^{d-1}x\ (\psi_{\rm spherical}^{(E,\lambda)}(x))^{\dagger}\psi_{\rm plane}^{(E',\hat{k},s)}(x) & = & \int d^{d-1}x\sum_{\lambda'}(\psi_{\rm spherical}^{(E,\lambda)}(x))^{\dagger}\alpha(E',\hat{k},s,\lambda') \psi_{\rm spherical}^{(E',\lambda')}(x) \nonumber\\
& = & 2 \pi \delta(E-E') \alpha(E',\hat{k},s,\lambda) \,\,\, .
\end{eqnarray}
Similarly:
\begin{eqnarray}
\int d^{d-1}x'\ (\psi_{\rm plane}^{(E',\hat{k},s)}(x'))^{\dagger}\psi_{\rm spherical}^{(E'',\lambda)}(x') & = & \int d^{d-1}x'\sum_{\lambda'}(\psi_{\rm spherical}^{(E',\lambda')}(x'))^{\dagger}\alpha^{*}(E',\hat{k},s,\lambda') \psi_{\rm spherical}^{(E'',\lambda)}(x') \nonumber\\
& = & 2 \pi \delta(E'-E'')\alpha^{*}(E',\hat{k},s,\lambda) \,\,\, .
\end{eqnarray}
Multiplying these two expressions by $E'^{d-2}$, summing over $s$, and integrating over $dE' d\Omega_{d-2}^{(\hat{k})}$, we have:
\begin{eqnarray}
\sum_{s=\pm} \int dE'E'^{d-2}d\Omega_{d-2}^{(\hat{k})}(2\pi)^{2}\delta(E-E')\delta(E'-E'')
|\alpha(E',\hat{k},s,\lambda)|^{2} \nonumber \\
\hspace{2 cm} = (2\pi)^{2}\delta(E-E'')E^{d-2}\sum_{s=\pm}\int d\Omega_{d-2}^{(\hat{k})}|\alpha(E,\hat{k},s,\lambda)|^{2} \,\,\, . \nonumber\\
\label{final}
\end{eqnarray}
We can also evaluate this expression another way:
\begin{eqnarray}
\sum_{s=\pm}\int & dE' & E'^{d-2}\int d\Omega_{d-2}^{(\hat{k})}\int d^{d-1}x\int d^{d-1}x'(\psi_{\rm spherical}^{(E,\lambda)}(x))^{\dagger}\psi_{\rm plane}^{(E',\hat{k},s)}(x) (\psi_{\rm plane}^{(E',\hat{k},s)}(x'))^{\dagger}\psi_{\rm spherical}^{(E'',\lambda)}(x') \nonumber\\
& = & \int d^{d-1}x\int d^{d-1}x'(\psi_{\rm spherical}^{(E,\lambda)}(x))^{\dagger} \left( \sum_{s=\pm}\int dE'E'^{d-2}\int d\Omega_{d-2}^{(\hat{k})}\psi_{\rm plane}^{(E',\hat{k},s)}(x)(\psi_{\rm plane}^{(E',\hat{k},s)}(x'))^{\dagger}\right)\psi_{\rm spherical}^{(E'',\lambda)}(x') \,\,\, . \nonumber\\  \label{projection}
\end{eqnarray}
For the expression in the bracket:
\begin{eqnarray}
\sum_{s=\pm}\int dE'E'^{d-2} \int d\Omega_{d-2}^{(\hat{k})}\psi_{\rm plane}^{(E',\hat{k},s)}(x)(\psi_{\rm plane}^{(E',\hat{k},s)}(x'))^{\dagger} = \int d^{d-1}k e^{i\vec{k}\cdot(\vec{x}-\vec{x}')} \sum_{s=\pm}\psi_{0}^{(\hat{k},s)}(\psi_{0}^{\hat{k},s)})^{\dagger} \,\,\, , \label{planewave}
\end{eqnarray}
and the sum over $s$ (of the constant spinors) is actually a projection operator. From equation (\ref{constspinor}) we have:
\begin{eqnarray}
\frac{1}{2}(1 + \hat{k}_{i}\gamma^{0}\gamma^{i})\psi_{0}^{(\hat{k},s)} = 0 & \Rightarrow & \frac{1}{2}(1 + \hat{k}_{i}\gamma^{0}\gamma^{i}) \sum_{s=\pm}\psi_{0}^{(\hat{k},s)}(\psi_{0}^{\hat{k},s)})^{\dagger} = 0 \,\,\, ,
\end{eqnarray}
hence:
\begin{equation}
\sum_{\pm}\psi_{0}^{(\hat{k},s)} (\psi_{0}^{\hat{k},s)})^{\dagger} = \frac{1}{2}(1-\hat{k}_{i}\gamma^{0}\gamma^{i})\equiv\Lambda_{+} \,\,\, ,
\end{equation}
where we have defined the projection operators for positive and negative energy solutions as:
\begin{equation}
\Lambda_{\pm}\equiv\frac{1}{2}(1\mp\hat{k}_{i}\gamma^{0}\gamma^{i}) \,\,\, ,
\end{equation}
with $\Lambda_{+}^{2}=\Lambda_{+}$, $\Lambda_{-}^{2}=\Lambda_{-}$, $\Lambda_{+} \Lambda_{-} = \Lambda_{-} \Lambda_{+} = 0$, and $\Lambda_{+}+\Lambda_{-}=1$. With this result we can write equation (\ref{planewave}) as:
\begin{eqnarray}
\int d^{d-1}k\ \frac{1}{2}(1-\hat{k}_{i}\gamma^{0}\gamma^{i})
e^{i\vec{k}\cdot(\vec{x}-\vec{x}')}&=&\frac{1}{2}
\left(1+\frac{i}{\sqrt{-\vec{\partial}^{2}}}
\gamma^{0}\gamma^{i}\partial_{i}\right)\int d^{d-1}k\
e^{i\vec{k}\cdot(\vec{x}-\vec{x}')}\nonumber\\
&=&\frac{1}{2}\left(1+\frac{i}{\sqrt{-\vec{\partial}^{2}}}
\gamma^{0}\gamma^{i}\partial_{i}\right)(2\pi)^{d-1}\delta(\vec{x}-\vec{x}') \,\,\, .
\end{eqnarray}
Finally, returning to equation (\ref{projection}):
\begin{eqnarray}
\int d^{d-1}x\int d^{d-1}x'(\psi_{\rm
spherical}^{(E,\lambda)}(x))^{\dagger}\frac{1}{2}
\left(1+\frac{i}{\sqrt{-\vec{\partial}^{2}}}
\gamma^{0}\gamma^{i}\partial_{i}\right)
(2\pi)^{d-1}\delta(\vec{x}-\vec{x}')\psi_{\rm
spherical}^{(E'',\lambda)}(x')\nonumber\\
\hspace{1in}=(2\pi)^{d-1}\int d^{d-1}x(\psi_{\rm
spherical}^{(E,\lambda)}(x))^{\dagger}\frac{1}{2}
\left(1+\frac{i}{\sqrt{-\vec{\partial}^{2}}}
\gamma^{0}\gamma^{i}\partial_{i}\right)\psi_{\rm
spherical}^{(E'',\lambda)}(x)\label{spherical} \,\,\, .
\end{eqnarray}
Since the spherical spinor waves also have positive energies, they satisfy:
\begin{eqnarray}
\gamma^{\mu}\partial_{\mu}\psi_{\rm spherical}^{(E'',\lambda)}=0
&\Rightarrow&\gamma^{i}\partial_{i}\psi_{\rm
spherical}^{(E'',\lambda)}=iE''\gamma^{0}\psi_{\rm
spherical}^{(E'',\lambda)}\nonumber\\
&\Rightarrow&-\vec{\partial}^{2}\psi_{\rm
spherical}^{(E'',\lambda)}=E''^{2}\psi_{\rm
spherical}^{(E'',\lambda)} \,\,\, .
\end{eqnarray}
In which case equation (\ref{spherical}) becomes:
\begin{eqnarray}
(2\pi)^{d-1}\int d^{d-1}x(\psi_{\rm
spherical}^{(E,\lambda)}(x))^{\dagger} \psi_{\rm
spherical}^{(E'',\lambda)}(x)=(2\pi)^{d}\delta(E-E'') \,\,\, .
\end{eqnarray}

\par Equating this result to equation (\ref{final}) we have:
\begin{eqnarray}
\sum_{s=\pm}\int d\Omega_{d-2}^{(\hat{k})}|\alpha(E,\hat{k},s,\lambda)|^{2}=\left(
\frac{2\pi}{E}\right)^{d-2} \,\,\, ,
\end{eqnarray}
with the result that the cross-section in equation (\ref{sigma}) can be expressed as:
\begin{eqnarray}
\sigma=\frac{1}{2\Omega_{d-2}}\left(\frac{2\pi}{E}\right)^{d-2}
\sum_{\lambda}|{\cal A}^{s=1/2}(E,\lambda)|^{2}\label{finalsigma} \,\,\, .
\end{eqnarray}
To express the cross-section more explicitly, we first note that
the total volume of a unit $(d-2)$-sphere is:
\begin{eqnarray}
\Omega_{d-2}=\frac{2\pi^{(d-1)/2}}{\Gamma((d-1)/2)} \,\,\, ,
\end{eqnarray}
and that the Dirac operator eigenvalues of the eigenspinors on a
$(d-2)$-sphere are \cite{Camporesi}:
\begin{eqnarray}
\lambda_{l}=\pm i\left(\ell+\frac{d-2}{2}\right) \,\,\, ,
\end{eqnarray}
where $\ell=0,1,2,\dots$.  Note that the degeneracies of the eigenvalues, which are equal to the dimension of the spinor representations $SP(d-2)$ \cite{Camporesi}, are:
\begin{eqnarray}
D^{s=1/2}_{\ell}(d-2)=
\frac{2^{(d-2)/2}\Gamma(\ell+d-2)}{\Gamma(\ell+1)\Gamma(d-2)} \,\,\, ,
\end{eqnarray}
for even spheres, and
\begin{eqnarray}
D^{s=1/2}_{\ell}(d-2)=
\frac{2^{(d-3)/2}\Gamma(\ell+d-2)}{\Gamma(\ell+1)\Gamma(d-2)} \,\,\, ,
\end{eqnarray}
for odd spheres. One could write for even and odd spheres that:
\begin{eqnarray}
D^{s=1/2}_{\ell}(d-2)=
\frac{2^{[(d-2)/2]}\Gamma(\ell+d-2)}{\Gamma(\ell+1)\Gamma(d-2)} \,\,\, ,
\end{eqnarray}
where $[n]$ is the integral part of $n$.

\par Instead of $\ell$ we have previously used another parameter, $\kappa$, where:
\begin{eqnarray}
\kappa=\pm\left(\ell+\frac{d-2}{2}\right)=\pm\left(\frac{d}{2}-1\right),
\pm\frac{d}{2},\pm\left(\frac{d}{2}+1\right),\dots
\end{eqnarray}
or
\begin{eqnarray}
\ell=|\kappa|-\frac{d}{2}+1 \,\,\, .
\end{eqnarray}
Therefore:
\begin{eqnarray}
D_{\kappa}^{s=1/2}(d-2)=\frac{2^{[(d-2)/2]}\Gamma(|\kappa|+\frac{d}{2}-1)}
{\Gamma(|\kappa|-\frac{d}{2}+2)\Gamma(d-2)} \,\,\, . \label{degeneracy}
\end{eqnarray}
Finally, putting all these into equation (\ref{finalsigma}) we have:
\begin{eqnarray}
\sigma&=&\frac{1}{2\Omega_{d-2}}\left(\frac{2\pi}{E}\right)^{d-2}
\sum_{\kappa}D_{\kappa}^{s=1/2}(d-2)|{\cal
A}^{s=1/2}(E,\kappa)|^{2}\nonumber\\ &=&\frac{2^{[(d-6)/2]}(d-2)}
{\Gamma(d/2)}\left(\frac{\sqrt{\pi}}{E}\right)^{d-2}
\sum_{\kappa}\frac{\Gamma(|\kappa|+\frac{d}{2}-1)}
{\Gamma(|\kappa|-\frac{d}{2}+2)}|{\cal A}^{s=1/2}(E,\kappa)|^{2} \,\,\, ,
\end{eqnarray}
where
$\kappa=\pm(\frac{d}{2}-1),\pm\frac{d}{2},\pm(\frac{d}{2}+1),\dots$.
For $d=4$, we have:
\begin{eqnarray}
\sigma=\frac{\pi}{E^{2}}\sum_{\kappa}|\kappa||{\cal
A}^{s=1/2}(E,\kappa)|^{2} \,\,\, ,
\end{eqnarray}
where $\kappa=\pm 1,\pm 2,\dots$. Note that this is the same as the result
found by Unruh \cite{Unruh}.



\begin{thebibliography}{9}


\bibitem{ADD}
  N.~Arkani-Hamed, S.~Dimopoulos and G.~R.~Dvali,
  Phys.\ Lett.\ B {\bf 429}, 263 (1998).

\bibitem{BHacc}
  P.~C.~Argyres, S.~Dimopoulos and J.~March-Russell,
  Phys.\ Lett.\ B {\bf 441}, 96 (1998);
  S.~Dimopoulos and G.~Landsberg,
  Phys.\ Rev.\ Lett.\  {\bf 87}, 161602 (2001);
S.~B.~Giddings and S.~D.~Thomas,
  Phys.\ Rev.\ D {\bf 65}, 056010 (2002).


\bibitem{Split}
  N.~Arkani-Hamed and M.~Schmaltz,
  Phys.\ Rev.\ D {\bf 61}, 033005 (2000);
  N.~Arkani-Hamed, Y.~Grossman and M.~Schmaltz,
  Phys.\ Rev.\ D {\bf 61}, 115004 (2000);
  T.~Han, G.~D.~Kribs and B.~McElrath,
  Phys.\ Rev.\ Lett.\  {\bf 90}, 031601 (2003);

\bibitem{HiHawk}
P.~Kanti and J.~March-Russell,
{ Phys.\ Rev.}\ D {\bf 66}, 024023 (2002) [hep-ph/0203223];
V.~P.~Frolov and D.~Stojkovic,
{ Phys.\ Rev.}\ D {\bf 66}, 084002 (2002) [hep-th/0206046];
E.~l.~Jung, S.~H.~Kim and D.~K.~Park,
{ Phys.\ Lett.} B {\bf 586} (2004) 390 [hep-th/0311036];
E.~Jung and D.~K.~Park,
{ Nucl.\ Phys.}\ B {\bf 717} (2005) 272 [hep-th/0502002];
 D.~K.~Park,
  {Class.\ Quant.\ Grav.}  {\bf 23} (2006) 4101 [hep-th/0512021];
  V.~Cardoso, M.~Cavaglia and L.~Gualtieri,
  { Phys.\ Rev.\ Lett.}  {\bf 96}, 071301 (2006)
  [Erratum-ibid.\  {\bf 96}, 219902 (2006)] [hep-th/0512002];
  S.~Creek, O.~Efthimiou, P.~Kanti and K.~Tamvakis,
  { Phys.\ Lett.} B {\bf 635} (2006) 39 [hep-th/0601126];
  O.~Efthimiou,
  hep-th/0609144.

 \bibitem{Kanti}
  C.~M.~Harris and P.~Kanti,
{ JHEP} {\bf 0310}, 014 (2003) [hep-ph/0309054];
  P.~Kanti,
  Int.\ J.\ Mod.\ Phys.\  A {\bf 19} (2004) 4899
  [arXiv:hep-ph/0402168].

\bibitem{Rotate}
V.~P.~Frolov and D.~Stojkovic, 
{\it Phys.\ Rev.}\ D {\bf 67}, 084004 (2003) [gr-qc/0211055];
E.~Jung and D.~K.~Park, 
hep-th/0506204;
H.~Nomura, S.~Yoshida, M.~Tanabe and K.~i.~Maeda,
{ Prog.\ Theor.\ Phys.}  {\bf 114}, 707 (2005) [hep-th/0502179];
C.~M.~Harris and P.~Kanti,
{ Phys.\ Lett.} B {\bf 633}, 106 (2006) [hep-th/0503010];
D.~Ida, K.~y.~Oda and S.~C.~Park,
{ Phys.\ Rev.} D {\bf 71}, 124039 (2005) [hep-th/0503052];
{ Phys.\ Rev.} D {\bf 73}, 124022 (2006) [hep-th/0602188];
  G.~Duffy, C.~Harris, P.~Kanti and E.~Winstanley,
{ JHEP} {\bf {0509}}, 049 (2005) [hep-th/0507274];
M.~Casals, P.~Kanti and E.~Winstanley,
{JHEP} {\bf 0602}, 051 (2006) [hep-th/0511163];
S.~Creek, O.~Efthimiou, P.~Kanti and K.~Tamvakis,
  arXiv:hep-th/0701288.

 \bibitem{BraneFerm}
D.~Ida, K.~y.~Oda and S.~C.~Park,
{Phys.\ Rev.} D {\bf 67}, 064025 (2003)
[Erratum-ibid.\ D {\bf 69}, 049901 (2004)] [hep-th/0212108];
P.~Kanti and J.~March-Russell, 
{Phys.\ Rev.}\ D {\bf 67}, 104019 (2003) [hep-ph/0212199];
E.~l.~Jung, S.~H.~Kim and D.~K.~Park,
{Phys.\ Lett.} B {\bf 614} (2005) 78 [hep-th/0503027];
M.~Casals, S.~R.~Dolan, P.~Kanti and E.~Winstanley, hep-th/0608193.


\bibitem{CCDN}
  H.~T.~Cho, A.~S.~Cornell, J.~Doukas and W.~Naylor,
  {Phys.\ Rev.} D {\bf 75} (2007) 104005 [arXiv:hep-th/0701193].

\bibitem{Das}
  S.~R.~Das, G.~W.~Gibbons and S.~D.~Mathur,
  Phys.\ Rev.\ Lett.\  {\bf 78} (1997) 417
  [arXiv:hep-th/9609052].

\bibitem{Dai}
  D.~C.~Dai, N.~Kaloper, G.~D.~Starkman and D.~Stojkovic,
  [arXiv:hep-th/0611184];
N.~Kaloper and D.~Kiley,
  JHEP {\bf 0603}, 077 (2006)
  [arXiv:hep-th/0601110].
 S.~Chen, B.~Wang and R.~K.~Su,
  Phys.\ Lett.\  B {\bf 647}, 282 (2007)
  [arXiv:hep-th/0701209].
 U.~A.~al-Binni and G.~Siopsis,
  arXiv:0708.3363 [hep-th].


 \bibitem{Cornell}
  A.~S.~Cornell, W.~Naylor and M.~Sasaki,
  JHEP {\bf 0602} (2006) 012
  [arXiv:hep-th/0510009].

\bibitem{ChoWKB}
  H.~T.~Cho and Y.~C.~Lin,
  Class.\ Quant.\ Grav.\  {\bf 22} (2005) 775
  [arXiv:gr-qc/0411090].

\bibitem{Unruh} W.G. Unruh, Phys. Rev. {\bf D 14}, 3251 (1976).


\bibitem{JP} E.~l.~Jung, S.~H.~Kim and D.~K.~Park, 
{\it JHEP} {\bf 0409} (2004) 005 [hep-th/0406117];


\bibitem{IW}
S.~Iyer and C.~M.~Will,
Phys.\ Rev.\ D {\bf 35} (1987) 3621.

 \bibitem{WG}
 C.~M.~Will and J.~W.~Guinn, Phys.\ Rev.\ A {\bf 37} (1988) 3674.

 \bibitem{Jung}
  E.~Jung and D.~K.~Park,
  Nucl.\ Phys.\  B {\bf 766}, 269 (2007)
  [arXiv:hep-th/0610089].

  \bibitem{EHM}
  R.~Emparan, G.~T.~Horowitz and R.~C.~Myers,
  Phys.\ Rev.\ Lett.\  {\bf 85}, 499 (2000)
  [arXiv:hep-th/0003118].

\bibitem{Cardoso}
  V.~Cardoso, M.~Cavaglia and L.~Gualtieri,
  JHEP {\bf 0602}, 021 (2006)
  [arXiv:hep-th/0512116].

  \bibitem{LWY}
  L.~Liu, B.~Wang and G.~Yang,
  arXiv:hep-th/0701166.

  \bibitem{Camporesi}
  R.~Camporesi and A.~Higuchi,
  J.\ Geom.\ Phys.\  {\bf 20} (1996) 1
  [arXiv:gr-qc/9505009].

  \bibitem{Konoplya6th}
  R.~A.~Konoplya,
  Phys.\ Rev.\  D {\bf 68}, 024018 (2003)
  [arXiv:gr-qc/0303052].
 \bibitem{Kanti2}
S.~Creek, O.~Efthimiou, P.~Kanti and K.~Tamvakis,
  arXiv:0709.0241 [hep-th].
\end{thebibliography}
\end{document}